\documentclass[12pt]{article}
\usepackage{amsmath,amssymb,amscd,cite}
\newtheorem{thm}{Theorem}[section]
\newtheorem{prop}[thm]{Proposition}
\newtheorem{lem}[thm]{Lemma}
\newtheorem{cor}[thm]{Corollary}

\newtheorem{defi}[thm]{Definition}
\newcommand{\pf}{{\bf Proof. \ }}
\newcommand{\qed}{\hfill $\Box$ \\}
\font\msbm=msbm10 at 12pt
\newcommand{\Z}{\mbox{\msbm Z}}

\newcommand{\F}{\mbox{\msbm F}}
\newtheorem{rem}[thm]{Remark}

\date{}
\begin{document}

\title{ Algebraic Quantum Synchronizable Codes}

\author{K. Guenda, G. G. La Guardia  and T. A. Gulliver\thanks{K. Guenda
is with the Faculty of Mathematics USTHB, University of Science and
Technology of Algiers, Algeria.
G. G. La Guardia is with Department of Mathematics and Statistics,
State University of Ponta Grossa (UEPG), 84030-900, Ponta Grossa, PR, Brazil.
T. A. Gulliver is with the
Department of Electrical and Computer Engineering, University of
Victoria, PO Box 1700, STN CSC, Victoria, BC, Canada V8W 2Y2
email: kguenda@usthb.dz,
agullive@ece.uvic.ca.}}

\maketitle

\begin{abstract}

In this paper, we construct quantum synchronizable codes (QSCs) based on the sum and intersection of cyclic codes.
Further, infinite families of QSCs are obtained from BCH and duadic codes.
Moreover, we show that the work of Fujiwara~\cite{fujiwara1} can be generalized to repeated root cyclic codes (RRCCs) such that QSCs are always obtained,
which is not the case with simple root cyclic codes.
The usefulness of this extension is illustrated via examples of infinite families of QSCs from repeated root duadic codes.
Finally, QSCs are constructed from the product of cyclic codes.

\end{abstract}
\section{Introduction}

The main goal of frame synchronization in communication systems is to ensure that information block boundaries can be correctly determined at the receiver.
To achieve this goal, numerous synchronization techniques have been developed for classical communication systems.
However, these techniques are not applicable to quantum communication systems since a qubit measurement typically destroys the quantum states and
thus also the corresponding quantum information.
To circumvent this problem, synchronization can be achieved using a classical system external to the quantum system, but
such a solution does not take advantage of the benefits that quantum processing can provide.

In a landmark paper~\cite{fujiwara1}, Fujiwara provided a framework for quantum block synchronization.
The approach is to employ codes, called quantum synchronizable codes (QSCs), which allow
the identification of codeword boundaries without destroying the quantum states.
This is achieved by determining how many qubits from proper alignment the system is should misalignment occur.
More precisely, an $(a_l, a_r)-[[n, k]]_2$ QSC is an $[[n, k]]_2$ code
that encodes $k$ logical qubits into a physical qubit, and corrects
misalignments of up to $a_l$ qubits to the left and up to $a_r$ qubits to the right.
These quantum codes may correct more phase errors than bit errors.
This is an advantage because, as shown by
Ioffe and M\'ezard~\cite{ioffe}, in physical systems the noise is
typically asymmetric so that bit errors occur less frequently than phase errors.
Thus, one can consider QSCs as asymmetric quantum codes.
% even though the construction techniques differ.

The initial work by Fujiwara was improved  in~\cite{fujiwara2} by making more extensive use of finite algebra to obtain block QSCs.
Several QSC constructions have recently been presented~\cite{fujiwara3,fujiwara4,Ge}.
These constructions employ BCH codes, cyclic codes related to finite geometries, punctured
Reed-Muller codes, and quadratic residue codes and duadic codes of length $p^n$.
Fujiwara and Vandendriessche~\cite{fujiwara3} noted that
``One of the main hurdles in the theoretical study of quantum
synchronizable codes is that it is quite difficult to find suitable
classical error-correcting codes because the required algebraic
constraints are very severe and difficult to analyze."
In this paper, quantum synchronizable codes are constructed based on the sum and intersection of cyclic codes.
Further, we construct infinite families of quantum synchronizable codes from BCH and duadic codes.
Moreover, the work of Fujiwara~\cite{fujiwara1} is generalized to repeated root cyclic codes (RRCCs) such that a QSC is always obtained,
which is not the case with simple root cyclic codes.
The usefulness of this extension is illustrated with examples of infinite families of QSCs from repeated root duadic codes.
Finally, we construct QSCs from the product of cyclic codes.

The remainder of this paper is organized as follows.
In Section~\ref{sec2}, some preliminary results and definitions are provided.
Section~\ref{sec3} presents several new constructions of QSCs.
More specifically, new families of QSCs are derived from the sum and intersection of cyclic codes,
and from BCH codes, duadic codes, and repeated root cyclic codes (RRCCs).
The construction of good QSCs given by Fujiwara is extended to RRCCs.
In Section 4, QSCs are constructed from the product of cyclic codes.

\section{Preliminary Results}\label{sec2}
Before presenting the constructions of QSCs, we recall some
preliminary results which will be used in the next section.

\begin{lem}\cite[Lemma 3.1]{Lidl:1997}\label{orderpoly}
Let $f(x) \in \F_q [x]$ be a polynomial of degree $m \geq 1$ with $f(0)\neq 0$.
Then there exists a positive integer $ e \leq q^{m} -1$ such that $f(x) | (x^{e}-1)$.
\end{lem}
From Lemma~\ref{orderpoly} the order of a nonzero polynomial is defined as follows.

\begin{defi}\cite[Definition 3.2]{Lidl:1997}
Let $f(x) \in \F_q [x]$ be a nonzero polynomial.
If $f(0)\neq 0$, the \emph{order} of $f(x)$, denoted by $\operatorname{ord}(f)$, is defined as the
smallest positive integer such that $f(x) | (x^{e}-1)$. If $f(0)=0$,
then $f(x)=x^{h}g(x)$ where $h \in \mathbb{N}$ and $g(x) \in \F_q [x]$ with $g(0)\neq 0$ are uniquely determined.
In this case, $\operatorname{ord}(f)$, is defined to be $\operatorname{ord}(g)$.
\end{defi}

The following results are well known.
\begin{lem}\cite[Theorem 4]{Macwilliams:1977}
Let $C$ be a cyclic code of length $n$ over $\F_q$ generated by $g(x)$.
Then the dual code $C^{\perp}$ of $C$ is generated by
$g^{\bot}(x)=\frac{x^n-1}{g^{*}(x)}$ where $g^{*}(x)=x^{\deg(g(x))}g(x^{-1})$.
\end{lem}

\begin{lem}
If $f(x), g(x), h(x) \in \F_{q}[x]$ such that $f(x)=g(x)h(x)$, then
$f^{*}(x)=g(x)^{*}h^{*}(x)$.
\end{lem}
\pf The result is obvious for constant polynomials.
Assume that $\deg(g(x))=m \geq 1$ and $\deg (h(x))=n \geq 1$.
Since $\deg(f(x))= m + n$, it follows that $f^{*}(x)=x^{m+n}f(1/x)= x^{m}g(1/x)x^{n}h(1/x)=g(x)^{*}h^{*}(x)$.
\qed

\begin{lem}\cite[Lemma 3.6]{Lidl:1997}
\label{lem:ord} Let $f(x) \in \F_q [x]$ with $f(0)\neq 0$ and $m$
be a positive integer. Then $f(x) | (x^{m}-1)$ if and only if
$\operatorname{ord}(f) | m$. Further, if the minimal polynomial
$M_1(x)$ divides $f(x)$ then $\operatorname{ord}(f) = m$.
\end{lem}

We recall the following result by Fujiwara~\cite{fujiwara1}.

\begin{thm}\cite[Theorem 1]{fujiwara1}
\label{thm:fuji} Let $C$ be a dual-containing $[n, k_1, d_1]$ cyclic
code and $D$ be a $C$-containing $[n, k_2, d_2]$ cyclic code
with $k_1 < k_2$. Then, for any pair of nonnegative integers $(a_l,
a_r)$ satisfying $a_l +a_r < k_2 - k_1$, there exists an $(a_l,
a_r)-[[n + a_l + a_r, 2k_1 - n]]$ QSC that corrects up to at least
$\lfloor \frac{d_1-1}{ 2} \rfloor$ phase errors and up to at least
$\lfloor \frac{d_2-1}{2} \rfloor $ bit errors.
\end{thm}

Theorem~\ref{thm:fuji} was improved in terms of synchronization capability as follows.

\begin{thm}\cite[Lemma 3]{fujiwara2}\label{thmfuji1}
Let $C$ be a dual-containing $[n, k_1, d_1]$ cyclic code and let $D$
be a $C$-containing $[n, k_2, d_2]$ cyclic code with $k_1 < k_2$.
Assume that $h(x)$ and $g(x)$ are the generator polynomials of $C$ and $D$, respectively.
Define the polynomial $f(x)$ of degree $k_2 - k_1$ such that $h(x)=f(x)g(x)$ over $\F_2[x]/(x^{n}-1)$.
Then for any pair of nonnegative integers $(a_l, a_r)$ satisfying $a_l +a_r <
\operatorname{ord}(f(x))$, there exists an $(a_l, a_r)-[[n + a_l +
a_r, 2k_1 - n]]$  QSC that corrects up to at least $\lfloor \frac{d_1-1}{ 2} \rfloor$ phase errors
and up to at least $\lfloor \frac{d_2-1}{2} \rfloor $ bit errors.
\end{thm}

\begin{rem} \hfill
\begin{itemize}
\item The quantum codes given in Theorem~\ref{thm:fuji} may correct more
phase errors than bit errors since the number of phase errors (resp. bit errors) is related to the minimum distance $d_1$ of
the dual-containing code $C$ (resp. to the minimum distance
$d_2$ of the $C$-containing code $D$).
As explained in Section 1, this is an advantage of QSCs.

\item Ioffe and M\'ezard~\cite{ioffe} showed that in physical systems the noise is typically asymmetric, i.e. bit errors
occur less frequently than phase errors.
Based on this fact, there has been significant interest in
constructing good asymmetric quantum codes
\cite{ezerman1,GG,sarvepali,laguardia:2012,laguardia:2013}.

\item The quantity $a_l + a_r$ in Theorem~\ref{thmfuji1} is called
the maximum tolerance magnitude of synchronization errors.
From Lemma~\ref{lem:ord}, this quantity is less than $m$ and
is maximal if the polynomial $h(x)$ in Theorem~\ref{thmfuji1} is
divisible by $M_1(x)$.
\end{itemize}
\end{rem}

\section{New Quantum Synchronizable Codes}\label{sec3}

In this section, we present several new constructions of quantum synchronizable codes (QSCs).
More precisely, we construct new families of QSCs from cyclic codes including duadic codes and BCH codes.

\subsection{Quantum Synchronizable Codes from Cyclic Codes}\label{subsec3.2}

We now present two constructions of QSCs from cyclic codes.
The first one is based on the sum code of cyclic codes and
the second is obtained by considering the intersection of cyclic codes.

\begin{thm}\label{sumcode}
Let $n \geq 3$ be an integer such that $\gcd(n, 2)=1$ and suppose
that $m={\operatorname{ord}}_{n}(2)$. Let $C_1$ be an $[n, k_1,
d_1]$ dual-containing cyclic code and let $C_2$ be an $[n, k_2,
d_2]$ $C_1$-containing cyclic code. Further, let $C_3$ be an $[n,
k_3, d_3]$ cyclic code and $C_4$ be an $[n, k_4, d_4]$
$C_3$-containing cyclic code such that $\deg(\gcd(g_2(x), g_4(x)))<
\deg(\gcd(g_1(x), g_3(x)))$, where $g_i(x)$ is the generator
polynomial of $C_i$, $i=1, 2, 3, 4$.
Then, for any pair of nonnegative integers $(a_l, a_r)$ satisfying
$a_l +a_r < \deg(\gcd(g_1(x), g_3(x)))-\deg(\gcd(g_2(x), g_4(x)))$,
there exists an $(a_l, a_r)-[[n + a_l + a_r, n-2\deg(\gcd(g_1(x),
g_3(x)))]]$ QSC that corrects up to at least $\lfloor \frac{d-1}{2}
\rfloor$ phase errors and up to at least $\lfloor \frac{d^{*}-1}{2}
\rfloor $ bit errors, where $d$ is the minimum distance of the code
$C_1 + C_3$ and $d^{*}$ is the minimum distance of the code $C_2 +
C_4$.
\end{thm}
\pf Since the codes $C_1$ and $C_3$ are cyclic, the sum code $C_1 + C_3 = \{c_1 + c_3 | c_1 \in C_1 \mbox{ and } c_3 \in C_3\}$ is also cyclic.
Since $C_1 + C_3$ is generated by the polynomial $g(x)=\gcd(g_1(x), g_3(x))$, it follows that $C_1 \subset C_1 + C_3$, so $(C_1 + C_3)^{\perp}\subset C_{1}^{\perp}$.
As $C_1$ is dual-containing, it follows that $(C_1 + C_3)^{\perp}\subset C_{1}^{\perp} \subset C_1 \subset C_1 + C_3$,
i.e. the sum code $C_1 + C_3$ is also a dual-containing cyclic code.

Let $g_{\diamond}(x)=\gcd(g_2(x), g_4(x))$.
As $g_2(x)|g_1(x)$ and $g_4(x)|g_3(x)$, it follows that $g_{\diamond}(x)|g(x)$, and hence the
inclusion $C_1 + C_3 \subset C_2 + C_4$ holds, where $C_2 + C_4$ is also a cyclic code.
From $\deg(g_{\diamond}(x))< \deg(g(x))$, it follows that $C_1 + C_3 \subsetneq C_2 + C_4$.
Applying Theorem~\ref{thm:fuji} to the codes $C_1 + C_3$ and $C_2 + C_4$,
a QSC is obtained with parameters $(a_l, a_r)-[[n + a_l + a_r, n-2\deg(\gcd(g_1(x), g_3(x)))]]$, where $a_l +a_r <
\deg(\gcd(g_1(x), g_3(x)))-\deg(\gcd(g_2(x), g_4(x)))$,
and corrects up to at least $\lfloor \frac{d-1}{2} \rfloor$ phase errors and up to at least $\lfloor \frac{d^{*}-1}{2} \rfloor $ bit errors.
\qed

\begin{thm}\label{intercode}
Let $n \geq 3$ be an integer such that $\gcd(n, 2)=1$ and suppose
that $m={\operatorname{ord}}_{n}(2)$. Let $C_1$ be an $[n, k_1, d_1]$ self-orthogonal cyclic code.
Further, let $C_2$ and $C_3$ be two cyclic codes with parameters $[n, k_2, d_2]$ and $[n, k_3, d_3]$,
respectively, such that $\{0\} \subsetneq C_{3}^{\perp}\subsetneq C_1 \cap C_2$.
Then for any pair of nonnegative integers $(a_l, a_r)$ satisfying $a_l +a_r < n -\deg
(g_3(x))-\deg(\mbox{lcm}(g_1(x), g_2(x)))$, there exists an
$(a_l, a_r)-[[n + a_l + a_r, 2\deg(\operatorname{lcm}(g_1(x),
g_2(x)))-n]]$ QSC that corrects up to at least $\lfloor
\frac{d-1}{2} \rfloor$ phase errors and up to at least $\lfloor
\frac{d_3 -1}{2} \rfloor $ bit errors, where $d$ is the minimum
distance of the code $(C_1 \cap C_2)^{\perp}$, and $g_i(x)$ is the
generator polynomial of $C_i$, $i=1,2,3$.
\end{thm}
\pf Since the codes $C_1$ and $C_2$ are cyclic, it follows that the
code $C_1 \cap C_2$ is cyclic.
Thus its dual code $(C_1 \cap C_2)^{\perp}$ is also cyclic.
As $C_1 \cap C_2 \subset C_1$, the inclusion $C_1^{\perp} \subset (C_1 \cap C_2)^{\perp}$ holds.
Since $C_1$ is self-orthogonal, then $C_1 \cap C_2 \subset C_1 \subset
C_1^{\perp}\subset(C_1 \cap C_2)^{\perp}$,
i.e. the code $C_1 \cap C_2$ is self-orthogonal.
Hence $(C_1 \cap C_2)^{\perp}$ is a dual-containing cyclic code.
As $C_{3}^{\perp}\subsetneq C_1 \cap C_2$, we know that $(C_1 \cap C_2)^{\perp} \subsetneq C_3$.
The dimension of the corresponding quantum code is $2\deg(\operatorname{lcm}(g_1(x), g_2(x)))-n$
and $a_l +a_r < n -\deg (g_3(x))-\deg(\operatorname{lcm}(g_1(x), g_2(x)))$.
Applying Theorem~\ref{thm:fuji} to the codes $(C_1 \cap
C_2)^{\perp}$ and $C_3$, for any pair of nonnegative integers $(a_l,
a_r)$ satisfying $a_l +a_r < n -\deg
(g_3(x))-\deg(\operatorname{lcm}(g_1(x), g_2(x)))$, there exists an $(a_l, a_r)-[[n + a_l + a_r, 2\deg(\operatorname{lcm}(g_1(x), g_2(x)))-n]]$
QSC that corrects up to at least $\lfloor \frac{d-1}{2} \rfloor$ phase errors and up to at least $\lfloor \frac{d_3 -1}{2} \rfloor $ bit errors.
\qed

\subsection{Quantum Synchronizable Codes from BCH Codes}\label{subsec3.1}

The class of BCH codes \cite{BCH-I:1960,BCH-II:1960} has been extensively employed in the construction of quantum codes.
In \cite{fujiwara2}, primitive BCH codes were used to construct quantum synchonizable codes.
In this section, quantum synchronizable codes are constructed from BCH codes that are not primitive.
First, we recall some basic concepts regarding BCH codes.

Let $\gcd(n,q) =1$. The $q$-cyclotomic coset ($q$-coset for short),
of $s$ modulo $n$ is defined as $C_{s} =\{s, sq, \ldots , sq^{m_{s}-1} \}$, where $sq^{m_{s}}\equiv s \bmod n$.
Let $\alpha$ be a primitive $n$th root of unity and $M_{i}(x)$ denote the minimal
polynomial of ${\alpha}^{i}$.
With this notation, the class of BCH codes, which are a subclass of cyclic codes, can be defined as follows.
\begin{defi} A cyclic code of length $n$ over $\F_{q}$ is a BCH code
with design distance $\delta$ if for some $b \geq 0$, $g(x)=\operatorname{lcm} \{ M_{b}(x), M_{b+1}(x), \ldots , M_{b+ \delta-2}(x)\}$.
The generator polynomial $g(x)$ of $C$ can be expressed in terms of its defining set $Z = C_{b}\cup C_{b+1}\cup \cdots \cup
C_{b+\delta -2}$ as $g(x) = \displaystyle{\prod}_{z \in Z}(x -{\alpha}^{z})$.
\end{defi}
It is well-known from the BCH bound that the minimum
distance of a BCH code is greater than or equal to its design distance $\delta$.

Consider the following two useful results.
\begin{prop}
\label{prop:Aly}\cite[Theorems 3 and 10]{aly} Let $n$ be a positive
integer such that $\gcd(n,2)=1$ and let $m={\operatorname{ord}}_n(2)$.
If $2 \le \delta \le \delta _{max} = \lfloor \kappa \rfloor$,
where $\kappa = \frac{n}{2^{m}-1} ( 2^{\lceil m/2 \rceil} -1)$, then
the narrow-sense $BCH(n,2,\delta)$ code contains its Euclidean dual
$BCH^{\bot}(n,2,\delta)$.
\end{prop}

\begin{lem}\cite[Lemmas 8 and 9]{aly}\label{rr}
Let $n\geq 1$ be an integer such that $\gcd(n, 2)=1$ and
${2}^{\lfloor m/2 \rfloor} < n \leq 2^{m}-1$, where $m=
{\operatorname{ord}}_{n}(2)$.

\begin{enumerate}
\item[ (i)] The $2$-coset ${\mathbb{C}}_{x}$ has cardinality $m$ for all $x$ in the range $1\leq x \leq n {2}^{\lceil m/2\rceil}/(2^{m}-1)$.

\item[ (ii)] If $x$ and $y$ are distinct integers in the range $1 \leq x,y \leq \min \{ \lfloor n{2}^{\lceil m/2 \rceil }/(2^{m}-1)-1 \rfloor, n-1 \}$
such that $x, y \not\equiv 0 \bmod 2$, then the $2$-cosets of $x$ and $y \bmod n$ are disjoint.
\end{enumerate}
\end{lem}

We now construct the new QSCs.

\begin{thm}\label{BCH}
Let $n \geq 3$ be an integer such that $\gcd(n, 2)=1$ and suppose
that ${2}^{\lfloor m/2 \rfloor} < n \leq 2^{m}-1$, where $m={\operatorname{ord}}_{n}(2)$.
Consider integers $a$ and $b$ such that
$1 \leq a < b < r=\min \{ \lfloor n{2}^{\lceil m/2 \rceil}/(2^{m}-1)-1 \rfloor, n-1, \lfloor \kappa \rfloor \}$,
where $\kappa = \frac{n}{2^{m}-1} ( 2^{\lceil m/2 \rceil} -1)$ and $a, b \not\equiv 0 \bmod 2$.
Then, for any pair of nonnegative
integers $(a_l, a_r)$ satisfying $a_l +a_r < m(t-u)$, there exists
an $(a_l, a_r)-[[n + a_l + a_r, n-2m(t+1)]]$ QSC that corrects up to at
least $\lfloor \frac{d-1}{2} \rfloor$ phase errors and up to at
least $\lfloor \frac{d^{*}-1}{2} \rfloor $ bit errors, where
$d\geq b+1$, $d^{*}\geq a+1$, $t=(b-1)/2$ and $u=(a-1)/2$.
\end{thm}
\pf Let $D$ be the binary narrow-sense BCH code of length $n$
generated by the product of the minimal polynomials
\begin{eqnarray*}
D=\langle M_{1}(x) M_{3}(x) \cdots M_{a}(x)\rangle,
\end{eqnarray*}
where $a= 2u + 1$ and $u\geq 0$ is an integer.
Further, let $C$ be the binary narrow-sense BCH code of length
$n$ generated by the product of the minimal polynomials
\begin{eqnarray*}
C=\langle M_{1}(x) M_{3}(x)\cdots M_{b}(x)\rangle,
\end{eqnarray*}
where $b=2t + 1$ and $t\geq 1$ is an integer.
It then follows that $C \subset D$, and by Proposition~\ref{prop:Aly} $C$ is dual-containing.
From Lemma~\ref{rr} and a straightforward
computation, the dimension of $D$ is $k_2 = n - m(u+1)$.
Similarly, the dimension of $C$ is $k_1 = n - m(t+1)$.
Thus, $k_2 - k_1 =m(t-u)$ and $2k_1 - n = n - 2m(t+1)$.
From the BCH bound, since the defining set of $D$ contains a sequence of $a$ consecutive integers, it
follows that the minimum distance of $D$ satisfies $d_2 \geq a+1$.
Analogously, since the defining set of $C$ contains a sequence
of $b$ consecutive integers, from the BCH bound the minimum
distance of $C$ satisfies $d_1 \geq b+1$.
The result then follows from Theorem~\ref{thm:fuji}. \qed

\begin{rem}
Let $f(x) \in \F_q [x]$.
Since $\operatorname{ord}(f(x)) \geq \deg (f(x))$, by applying Theorem~\ref{thmfuji1} one can
improve the upper bound for $a_l +a_r$, i.e. $a_l +a_r < m(t-u)\leq
\operatorname{ord}(M_{a+1}(x) \cdots M_{b}(x))$.
\end{rem}

We now construct QSCs from the sum of BCH codes.

\begin{thm}\label{sumBCH}
Let $n \geq 3$ be an integer such that $\gcd(n, 2)=1$ and suppose
that ${2}^{\lfloor m/2 \rfloor} < n \leq 2^{m}-1$, where
$m={\operatorname{ord}}_{n}(2)$. Consider integers $a$, $b$, $e$ and
$f$ such that $2 \leq e < a < b < f < \min \{ \lfloor n{2}^{\lceil
m/2 \rceil}/(2^{m}-1)-1 \rfloor, n-1, \lfloor \kappa \rfloor \}$,
where $\kappa = \frac{n}{2^{m}-1} ( 2^{\lceil m/2 \rceil} -1)$ and
$a, b, e, f \not\equiv 0 \bmod 2$.
Then, for any pair of nonnegative integers $(a_l, a_r)$ satisfying $a_l +a_r < m(t-w)$, there exists
an $(a_l, a_r)-[[n + a_l + a_r, n-2m(t+1)]]$ QSC that corrects up to
at least $\lfloor \frac{d-1}{2} \rfloor$ phase errors and up to at
least $\lfloor \frac{d^{*}-1}{2} \rfloor $ bit errors, where $d\geq
b+1$, $d^{*}\geq e+1$, $t=(b-1)/2$ and $w=(e-1)/2$.
\end{thm}

\pf Let $C_1$ be the binary narrow-sense BCH code of length $n$
generated by the product of the minimal polynomials
\begin{eqnarray*}
C_1=\langle M^{(1)}(x) M^{(3)}(x)\cdots M^{(b)}(x)\rangle,
\end{eqnarray*}
where $b=2t + 1$ and $t \geq 0$. Let $C_2$ be the binary
narrow-sense BCH code of length $n$ generated by the product of the
minimal polynomials
\begin{eqnarray*}
C_2=\langle M^{(1)}(x) M^{(3)}(x) \cdots M^{(a)}(x)\rangle,
\end{eqnarray*}
where $a= 2u + 1$ and $u\geq 1$.
From the construction $C_1 \subset C_2$, and by Proposition~\ref{prop:Aly} $C_1$ is dual-containing.
Further, consider the binary narrow-sense BCH codes of length $n$ generated by
\begin{eqnarray*}
C_3=\langle M^{(1)}(x) M^{(3)}(x)\cdots M^{(f)}(x)\rangle,
\end{eqnarray*}
and
\begin{eqnarray*}
C_4=\langle M^{(1)}(x) M^{(3)}(x) \cdots M^{(e)}(x)\rangle,
\end{eqnarray*}
where $f=2v + 1$ with $v\geq 1$, and $e=2w + 1$ with $w\geq 0$.
From the construction we have $C_3 \subsetneq C_4$. It then follows that $C_1
+ C_3 \subsetneq C_2 + C_4$. Since $e < a$ and $b < f$, from
Lemma~\ref{rr} and a straightforward computation, the codes $C_2 +
C_4$ and $C_1 + C_3$ have dimensions $K_2 = n - m(w+1)$ and $K_1 = n - m(t+1)$, respectively.
The dimension of the corresponding QSC is $K= n - 2m(t+1)$ and $K_2 - K_1 =  m (t - w)$.
Since $C_1$ is dual-containing, proceeding similar to the proof of
Theorem~\ref{sumcode}, it follows that $C_1 + C_3$ is also dual-containing.
From the BCH bound, the minimum distance $d_{13}$ of $C_1 + C_3$ satisfies $d_{13} \geq b+1$ and the minimum distance
$d_{24}$ of $C_2 + C_4$ satisfies $d_{24} \geq e+1$.
Applying Theorem~\ref{sumcode} to the codes $C_1 + C_3$ and $C_2 + C_4$, the result follows.
\qed

\subsection{Quantum Synchronizable Codes From Duadic Codes}\label{subsec3.3}

The duadic codes are a subclass of cyclic codes, and are a generalization of quadratic residue codes.
Smid~\cite{smid} characterized duadic codes based on the existence of a splitting.
Duadic codes are important because they are related to self-dual and isodual codes~\cite{G}.

As mentioned in Section 1, Zhang and Ge~\cite{Ge} constructed
algebraic synchronizable codes from duadic codes of length $p^n$.
Here we provide a more general result by considering duadic codes of length
$n$, where $n$ is a product of prime powers.
We first recall some results on duadic codes.

Let $S_1$ and $S_2$ be unions of $2$-cosets modulo $m$ such that
$S_1 \cap S_2 = \emptyset$, $S_1 \cup S_2 = \Z_m \setminus \{0\}$
and $\mu_aS_i \bmod m = S_{(i+1) \bmod 2}$.
The triple $\mu _a, S_1, S_2$ is called a splitting modulo $m$.
The odd-like duadic codes $D_1$ and $D_2$ are the cyclic codes over $\mathbb{F}_2$ with
defining sets $S_1$ and $S_2$, respectively, and generator
polynomials $ f_{1}(x)= \Pi_{i\in S_{1}}(x-\alpha^{i})$ and $f_{2}(x) \,=\,\Pi_{i\in S_{2}}(x-\alpha^{i})$, respectively.
The even-like duadic codes $C_1$ and $C_2$ are the cyclic codes over
$\mathbb{F}_2$ with defining sets $\{0\}\cup S_1$ and $\{0\} \cup S_2$, respectively.
The cardinality of $S_i$ is equal to $\frac{m-1}{2}$.
If the splitting is given by $\mu_{-1}$, then the minimum distance of the odd-like duadic codes satisfies $d^2-d+1 \ge m$.
This is known as the square root bound.

Let $m$ be an odd integer and denote the multiplicative order of $2$ modulo $m$ by ${\operatorname{ord}}_m(2)$.
This order is equal to the degree of the minimal polynomial $M_1(x)$, and is the smallest integer $l$ such that $2^l\equiv 1 \bmod m$.

In the following we give necessary and sufficient conditions for the existence of duadic codes.
The notation $x \; \square \; y$ means that $x$ is a quadratic residue modulo $y$.
\begin{thm}
\label{thm:smid1}\cite{smid} Duadic codes of length $m$ over $\F_2$ exist if and only if $2=\square \bmod m$.
In other words, if $m =p_1^{s_1}p_2^{s_2}\cdots p_k^{s_k}$ is the prime factorization of
$m$ where $s_i>0$, then duadic codes of length $m$ over $\F_2$ exist
if and only if $2=\square \bmod p_i$, $i=1,2,\ldots,k$.
\end{thm}

The following lemma shows that under certain conditions on
${\operatorname{ord}}_q(2)$, there is a specific factorization of $x^m-1$.

\begin{lem}\cite[Lemma 3.6]{GGISIT}
\label{lem:nontriv} Let $q$ be a prime power and $m$ be an odd
integer such that $\gcd(m, q)=1$, and suppose that ${\operatorname{ord}}_{m}(q)$ is odd.
Then any non-trivial irreducible divisor $M_i(x)$ of $x^{m}-1 $ in $\F_{q}[x]$ satisfies
$M_{i}(x)\neq \alpha M_{i}^*(x).\,\, \forall \alpha\in \F_{q}^*$.
\end{lem}

It can immediately be deduced from Lemma~\ref{lem:nontriv} that if $m$
is an odd integer such that ${\operatorname{ord}}_m (2)$ is odd,
then the polynomial $x^m-1$ can be decomposed as
\begin{equation}
\label{eq:decom} x^m-1=(x-1)M_{i_1}(x){M_{i_1}}^*(x) \cdots M_{i_s}(x){M_{i_s}}^*(x).
\end{equation}

We now investigate when a splitting modulo $m$, $m$ an odd integer, is given by the multiplier $\mu_{-1}$.

\begin{prop}
\label{prop:1} Let $m=p_{1}^{\alpha_1}\ldots  p_{l}^{\alpha_l}$ be
an odd integer such that, for all $i=1, \ldots, l$, $p_i \equiv -1 \bmod 8$.
Then the following hold:
\begin{itemize}
\item[(i)] all the splittings modulo $m$ are given by $\mu_{-1}$, and
\item[(ii)] there exists a pair of odd-like duadic codes $D_i$, $i=1,2$, generated by $g_i(x)$ such that $g_1(x)=g_2^*(x)$.
\end{itemize}
\end{prop}
\pf For part (i),
since $p_i\equiv -1 \bmod 8$ for all $i=1, \ldots, l$, it follows from~\cite[Theorem 8]{smid} that all the splitting are given by $\mu_{-1}$.
Part (ii) follows from part (i) and the decomposition in (\ref{eq:decom}).
\qed

\begin{prop}
\label{prop:dual}
Assume that $m={p_1}^{\alpha_1} \ldots {p_l}^{\alpha_l}$ is an odd integer such that, for all $i=1, \ldots, l$, $p_i\equiv -1 \bmod 8$.
Further assume that $T\subset \{i_1,\ldots,i_s\}$, where the $i_j$ are as given in (\ref{eq:decom}).
Consider the cyclic code $C$ with generator polynomial $g(x)= \prod_{j \in T} M_j(x)$.
Then $C$ is a dual-containing code.
\end{prop}
\pf Let $m={p_1}^{\alpha_1} \ldots {p_l}^{\alpha_l}$ be an odd
integer such that, for all $i=1, \ldots, l$, $p_i\equiv -1 \bmod 8$.
Then ${\operatorname{ord}}_{m} (2)= \mbox{lcm}({\operatorname{ord}}_{p_i}(2))$, which is odd, and
from~(\ref{eq:decom}), we have the decomposition
$x^m-1=(x-1)M_{i_1}(x){M_{i_1}}^{*}(x)\ldots M_{i_s}(x){M_{i_s}}^*(x)$.
Assume now that $g(x)= \prod_{i_j \in T} M_{i_j}(x)$ with $T=\{i_1,\ldots,i_t\}$, where $ t\le s$.
The dual code of $C$ has generator polynomial $g^{\bot}(x)=\frac{x^m-1}{\prod_{ i_1  \le i_j \le i_t} {M_{i_j}(x)}^*}$.
Hence, from (\ref{eq:decom}), we obtain that $g^{\bot}(x)=(x-1)\prod_{i_j \le t}M_{i_1}(x)\cdots M_{i_t}(x){M_{i_{t+1}}}^*(x)\cdots {M_{i_s}}^*(x)$, so $g|g^{\bot}$.
Therefore ${C}^{\bot} \subset C$ as required.
\qed

\begin{thm}
Let $m={p_1}^{\alpha_1} \ldots {p_1}^{\alpha_l}$ be an odd integer such that $p_i\equiv -1 \bmod 8$ for all $i=1, \ldots, l$.
Assume that $x^m-1$ can be decomposed as $x^m-1=(x-1)M_{i_1}(x)M_{i_1}^*(x)\cdots M_{i_s}(x)$
${M_{i_s}}^*(x)$, where the $M_{i_j}(x)$ are the minimal polynomials that are not self-reciprocal.
Further, assume that $T' \subset
T\subset \{i_1,\ldots,i_s\}$ are such that $\gcd(n; i_1; \ldots ; i_t)=1$ for all $i_j \in T$.
Then for any pair of non-negative integers $(a_l, a_r)$ satisfying $a_l+a_r< n$, with
$n=\operatorname{ord}(h(x))$, where $h(x)= \prod_{i_j\in T\setminus T'}M_{i_j}(x)$ and $n$ divides $m$,
there exists an $(a_l, a_r)-[[m+a_l+ a_r, m - 2\sum_{{i_j}\in T'} \deg (M_{i_j}(x))]]_2$ QSC that
corrects up to at least $\lfloor\frac{d_1-1}{2} \rfloor\ge 1$ phase errors and
up to at least $\lfloor \frac{d_2-1}{2} \rfloor \ge 1$ bit errors.
Here, $d_2 \le d_1$, where $d_1$ is the minimum distance of $C= \langle\prod_{i_j\in T} M_{i_j}(x)\rangle$ and $d_2$
is the minimum distance of $D=\langle\prod_{i_j\in T'}M_{i_j}(x)\rangle$.
\end{thm}
\pf Under the assumption ${\operatorname{ord}}_m(2)$, we obtain the
factorization of $x^m-1$ from (\ref{eq:decom}).
Assume that $T \subset \{ i_1,\ldots,i_s\}$, and let $g(x)=\prod_{i_j\in T} M_{i_j}(x)$.
Now consider the cyclic code $C$ generated by $g(x)$.
From Proposition~\ref{prop:dual} we have that ${C}^{\bot} \subset C$.

Let $D$ be the cyclic code generated by $h(x)=\prod_{{i_j}\in T'}M_{i_j}(x)$ with $T' \subset T$.
Then we have that $C \subset D$.
The polynomial $f(x)= \prod_{i_j\in T\setminus T'}M_{i_j}(x)$ has order $n$ a divisor of $m$.
From Lemma~\ref{lem:ord}, the dimension of $C$ is $k_1 = m-\sum_{i_j\in T} \deg (M_{i_j}(x))$,
and the dimension of $D$ is $k_2 = m-\sum_{i_j\in T'} \deg (M_{i_j}(x))$.
Hence from Theorem~\ref{thm:fuji}, for any pair of non-negative integers $(a_l, a_r)$ satisfying $a_l+a_r< n$,
there exists an $(a_l, a_r)-[[m+a_l+a_r, m- 2\sum_{i_j\in T'} \deg({M_{i_j}}(x))]]_2$ QSC
that corrects up to at least $\frac{d_1-1}{2}$ phase errors and up to at least $\frac{d_2-1}{2} $ bit errors.
The condition $\gcd(n; i_1; \ldots ; i_t)=1$ for all $i_j \in T$
ensures that the minimum distances $d_1$ and $d_2$ of the codes $C$
and $D$, respectively, are at least three \cite{charpin}.
\qed

\begin{cor}
\label{rem:2} Let $m={p_1}^{\alpha_1} \ldots {p_1}^{\alpha_l}$ be an odd integer such that
$p_i\equiv -1 \bmod 8$ for all $i=1, \ldots, l$.
Then there exists a QSC with parameters $(a_r,a_l)-[[m+a_r+a_l,1]]_2$ that can correct up to
$\lfloor \frac{d_1-1}{2} \rfloor$ phase errors, where $d_1^2-d_1+1 \ge m$, and
can correct $\lfloor \frac{d_2-1}{2}\rfloor$ bits errors, with $d_2 \ge \frac{m-1}{2}$.
\end{cor}
\pf Note that in this case we have $C= D_1=\langle\prod _{1 \le i_j \le \frac{m-1}{2}}M_{i_j}(x)\rangle$
and \ $D=\langle \prod _{1<i_j \le \frac{m-1}{2}}M_{i_j}(x)\rangle$.
If $f_1(x)$ and $g(x)$ are the generator polynomials of cyclic codes $C$ and $D$, respectively, then
$h(x)=f_1(x)/g(x)=M_1(x)$.
Hence from \cite[Theorem 3.5]{lidl}, the order of $h(x)$ equals $\operatorname{ord}_m(q)$.
The minimum distance of $C= D_2$ is equal to $d_1$, since the splitting is given by $\mu_{-1}$.
The computation of $d_1$ follows from the square root bound whereas the minimum distance $d_2$ of $D$ is
obtained from the BCH bound.
\qed

\begin{rem}
The codes given in Corollary~\ref{rem:2} can only encode one qubit.
Thus, even if the number of phase and bit errors which can be corrected is large, the code has very limited usefulness.
To avoid this situation and to take advantage of the previous construction,
repeated root cyclic codes (RRCCs) are considered in the next section to construct QSCs.
\end{rem}

\subsection{Quantum Synchronizable Codes from Repeated Root Cyclic Codes}\label{subsec3.4}
%The advantage of RRCCs is that they provide flexibility in the construction of QSCs.
In \cite{fujiwara3}, Fujiwara and Vandendriessche suggested that their results
may be generalized to lengths other than $2^m-1$.
However, it is difficult to determine the order of the generator polynomial in the repeated root case.
In this section, we show that the generalization of the construction of QSCs
to the repeated root cyclic code (RRCC) case can be done easily.
Further, employing RRCCs provides more flexibility and possibilities as a QSC can always be obtained.
%correction capability, especially if one has some requirement and
This generalization is possible due to the following remark.

\begin{rem}\hfill
\begin{itemize}
\item The properties of cyclic codes used for encoding and decoding in the synchronization scheme
suggested by Fujiwara et. al~\cite{fujiwara2} are that a cyclic shift of a codeword is also a codeword
and the polynomial representation of codewords.
Hence when considering the codes $C$ and $D$ such that $(C)^{\bot} \subset C \subset D$ as RRCCs,
the encoding and decoding does not have to change.

\item The maximal tolerable magnitude of synchronization errors is related to the order of the polynomial $f(x)$ as follows.
If $n=2^{n'}m$, $m$ odd, then we have $x^n-1=(x^m-1)^{2^{n'}}$.
Hence we obtain the factorization $x^n-1={f_1}^{2^{n'}} \ldots {f_l}^{2^{n'}}$.
If $f(x)$ is a power of an irreducible polynomial in
$\F_2[x]$ which is a divisor of $x^m-1$, then $f^{2^b}(x)| (x^{2^bm}-1)$.
From the definition one has
\begin{eqnarray}\label{ord:eq}
\operatorname{ord}( {f}^{2^b})= {(\operatorname{ord}(f))}^{2^b}.
\end{eqnarray}
Hence, the order of any divisor $f(x)$ of $x^n-1$ can be computed as
the least common multiple of the order of the power of irreducible
factors of $x^n-1$ as done previously.

\item From the previous remark, the order of the polynomial $h(x)$ used in the
construction of a code of length $2^am$ is at most equal to $2^a{\operatorname{ord}}(h'(x))$
where $h'(x)$ is the product of all irreducible polynomials which divide $h(x)$.
This makes the length of a QSC obtained from RRCCs larger by at most a factor $2^a$.
The next Theorem given by Castagnoli et al.~\cite{repeat} shows that while the length of the RRCC
increases, the minimum distance may also increase by a factor $P_t$.
Hence when using RRCCs to construct a QSC, we gain in the flexibility of choosing good codes in
Theorems~\ref{thm:fuji} and \ref{thmfuji1}, and also gain in error correcting capability without a loss in the error rate $d/n$.
\end{itemize}
\end{rem}

\begin{thm}\cite[Theorem 1]{repeat}
\label{thm:consta} Let $C$ be a $q$-ary repeated root cyclic code of
length $n=p^{\delta}m$, generated by $g(x)$, where $p$ is the
characteristic of $\F_q$, $\delta \ge 1$ and $\gcd(p,m)=1$.
Then $d_{min}(C)=P_t d_{min}(\overline{C_t})$ for some $t\in T$, where $\overline{C_t}$ is the cyclic code over $\F_q$ generated by
$\overline{g_t}$ (the product of the irreducible factors of $g(x)$ that occur with multiplicity $e_i >t$, and
$P_t=p^{j-1}(r+1)$, where $r$ is such that $t=(p-1)p^{\delta -1}+\ldots + (p-1)p^{\delta-(j-1)}+rp^{\delta -j}$.
\end{thm}

Next, we extend the construction presented in Section~\ref{sec2} to RRCCs.
We also show the importance of the flexibility in choosing codes with good minimum distance.
While it is difficult and challenging to find dual-containing
simple root cyclic codes, it is always possible to construct
dual-containing repeated root cyclic codes.
For instance, QSCs can always be constructed from cyclic codes of length $2n$, $n$ odd, with $x^n-1=f(x)g(x)$.
The dual of $f(x)$ is $f^{\bot}(x)=\frac{f(x)^{2*}g(x)^{2*}}{{f}^*(x)}$ and independent
of $f^*(x)$, we always have that $f(x)|f^{\bot}(x)$.
Hence the binary cyclic code of length $2n$ generated by $f(x)$ is dual-containing.
From~Theorem~\ref{thm:consta}, the minimum distance of this code is the same as the minimum distance of the code of length $n$.
If this minimum distance is good then there will not be a significant loss in the error rate.

Another interesting example is when the code length is $4n$.
In this case, if $x^n-1=f(x)g(x)$ it can easily be shown that the code $C$ of length $4n$ generated by $f^2(x)$ is dual-containing.
Further, we have the inclusion $C \subset D$, where $D$ is the cyclic code of length $4n$ generated by $f(x)$.
Hence if $M_1(x) |f(x)$, then we have that $\operatorname{ord}(f)=4n$.
Applying Theorems~\ref{thm:consta} and \ref{thm:fuji} gives the following result.
\begin{thm}
\label{thm:dua}
Let $n$ be an odd integer and assume that $x^n-1=f(x)g(x)$.
Then for any pair of nonnegative integers $(a_l, a_r)$ such that $a_l+a_r<4n$, there exists an
$(a_l,a_r)-[[4n, 4n-2 \deg (f(x))]]_2$ QSC that corrects at least $\frac{2d-1}{2}$ phase errors and
at least $\frac{d-1}{2}$ bit errors, where $d$ is the minimum distance of the cyclic code generated by $f(x)$.
\end{thm}

\begin{table}[h!]
\begin{center}
\caption{Some examples of  QSC obtained from Theorem~\ref{thm:dua}}
\label{table1}
\begin{tabular}{|c|c|c|c|c|c|}
\hline
  Linear Codes &Degree of $f$&QSC                                &Phase error                              &Bit error                              \\
\hline
$C=[n,k,d]_{2}$ &$d^{\circ} f$ &$Q_{s}=(a_l,a_r)-[[4*n,4*n-2d^{\circ} f]]_{2}$ &$\left\lfloor\frac{2d-1}{2}\right\rfloor$ &$\left\lfloor\frac{d-1}{2}\right\rfloor$ \\

\hline
$[7,4,3]_{2}$   &$3$           &$(20,5)-[[28,22]]_{2}$                         &$2$                                       &$1$                                       \\
\hline
$[5,1,5]_{2}$   &$4$           &$(2,5)-[[20,12]]_{2}$                          &$4$                                       &$2$                                       \\
\hline
$[17,9,5]_{2}$  &$8$           &$(30,8)-[[68,52]]_{2}$                         &$4$                                       &$2$                                       \\
\hline
$[19,1,19]_{2}$ &$18$           &$(31,13)-[[76,40]]_{2}$                        &$18$                                       &$9$                                       \\

\hline
$[27,9,3]_{2}$   &$18$           &$(70,15)-[[108,72]]_{2}$                         &$2$                                       &$1$                                       \\
\hline
$[47,24,11]_{2}$   &$23$           &$(57,113)-[[188,142]]_{2}$                          &$10$                                       &$5$                                       \\
\hline
$[71,36,11]_{2}$  &$35$           &$(150,23)-[[284,214]]_{2}$                         &$10$                                       &$5$                                       \\
\hline
$[97,49,15]_{2}$ &$48$           &$(103,215)-[[388,292]]_{2}$                        &$14$                                       &$7$                                       \\
\hline
$[103,52,19]_{2}$ &$51$           &$(250,91)-[[412,310]]_{2}$                        &$18$                                       &$9$                                       \\
\hline
\end{tabular}
\end{center}
\end{table}

\begin{rem}
If the splitting is given by $\mu_{-1}$, then the corresponding
duadic codes have good minimum distance.
The following theorem is an extension of the construction given in Section~\ref{sec2}.
\end{rem}

\begin{thm}
\label{thm: ordCons} Let $m={p_1}^{\alpha_1} \ldots {p_1}^{\alpha_l}$ be an odd integer such that,
for all $i=1, \ldots, l$, $p_i\equiv -1 \bmod 8$.
Then for any pair of nonnegative integers $(a_l, a_r)$ such that $a_l+a_r <2m$, there exists an
$(a_l,a_r)-[[2m, m-1]]_2$ $QSC$ that corrects at least $\frac{d-1}{2}$
bit and phase errors, where $d^2-d+1 \ge m$.
\end{thm}
\pf Let $m={p_1}^{\alpha_1} \ldots {p_1}^{\alpha_l}$ be an
odd integer such that $p_i\equiv -1 \bmod 8$ for all $i=1, \ldots,
l$. Then there exists a pair of odd-like duadic codes $D_i$, $i=1,2$, whose splitting is given by $\mu_{-1}$.
If $f_i(x)$ is the generator polynomial of $D_i$, then $x^m-1=(x-1)f_1(x)f_2(x)$, where ${f_1}^*(x)=f_2(x)$.
Let $C$ be the cyclic code of length $2m$ generated by $f(x)=(x-1)f_1(x)$.
The dimension of $C$ is $\frac{3m-1}{2}$.
Further, the polynomial $f^{\bot}(x)=\frac{(x-1)^2{f_1}^2{f_2}^2}{(x-1){f_1}^*}={f_1}^2f_2$ generates the dual code ${C}^{\bot}$.
If $D$ is the cyclic code of length $2m$ generated by $f_1(x)$, it follows that ${C}^{\bot} \subset C \subset D$.
From~(\ref{ord:eq}) and Lemma~\ref{lem:ord}, one has that $\operatorname{ord}f(x)=2m$.
Since the splitting is given by $\mu_{-1}$, the minimum distance of the duadic
code is $d$, and this is also the minimum distance of the cyclic
even-like duadic code of length $m$ generated by $(x-1)f_1(x)$.
From the square root bound it follows that $d^2-d+1 \ge m$.
Further, Theorem~\ref{thm:consta} gives that $d$ is also the minimum distance of the
codes $C$ and $D$.
\qed

The construction and proof of Theorem \ref{thm: ordCons} are valid when considering codes of length $2^im$, which gives the following theorem.

\begin{thm}
\label{thm: ordCon2} Let $m={p_1}^{\alpha_1} \ldots {p_l}^{\alpha_l}$ be an odd integer such that, for all $i=1, \ldots, l$, $p_i\equiv -1 \bmod 8$.
Then, for any pair $(a_l, a_r)$ of nonnegative integers such that $a_l + a_r < 2^im$, there exists an
$(a_l, a_r)-[[2^im, 2^im-m-1]]_2$ QSC that corrects at least
$\frac{d-1}{2}$ bit and phase errors, where $d^2-d+1 \ge m$.
\end{thm}

\section{Quantum Synchronizable Codes from Product Codes}\label{subsec3.6}
The product code construction is a useful means of combining codes of different length.
then in some cases the severe requirement on the cyclic codes (for example duadic and BCH codes) can be relaxed.
We recall the direct product of linear codes.
For more details we refer the reader to \cite{Macwilliams:1977}.

Let $C_1$ and $C_2$ be two linear codes with parameters $[n_1, k_1, d_1]_{q}$ and $[n_2, k_2, d_2]_{q}$, respectively, both over $\F_{q}$.
Assume that $G^{(1)}$ and $G^{(2)}$ are the generator matrices of $C_1$ and $C_2$, respectively.
Then the product code $C_1\otimes
C_2$ is a linear $[n_1 n_2, k_1 k_2, d_1 d_2]$ code over $\F_{q}$
generated by the Kronecker product matrix $G^{(1)} \otimes G^{(2)}$
defined as

\begin{eqnarray*}
G^{(1)} \otimes G^{(2)} = \left[
\begin{array}{cccc}
g_{11}^{(1)}G^{(2)} & g_{12}^{(1)}G^{(2)} &  \cdots & g_{1 n_1}^{(1)}G^{(2)}\\
g_{21}^{(1)}G^{(2)} & g_{22}^{(1)}G^{(2)} & \cdots & g_{2 n_1}^{(1)}G^{(2)}\\
\vdots & \vdots & \vdots & \vdots\\
g_{k_1 1}^{(1)}G^{(2)} & g_{k_1 2}^{(1)}G^{(2)} & \cdots & g_{k_1 n_1}^{(1)}G^{(2)}\\
\end{array}
\right]
\end{eqnarray*}

%In the next result we assume that all cyclic codes considered are binary.

\begin{thm}
Let $n$ and $n^{*}$ be two positive odd integers such that $\gcd( n,
n^{*})=1$. Let $C_1$ be an $[n, k_1, d_1]$ self-orthogonal cyclic
code and $C_2$ an $[n, k_2, d_2]$ cyclic code, both over $\F_2$.
Consider that $C_3$ and $C_4$ are two cyclic codes with parameters $[n^{*}, k_3, d_3]$
and $[n^{*}, k_4, d_4]$, respectively, over $\F_2$ such that ${(C_1 \otimes C_3 )}^{\perp} \subsetneq C_2 \otimes C_4$.
Then for any pair of nonnegative integers $(a_l, a_r)$ satisfying $a_l +a_r < k_1 k_3 + k_2 k_4 -nn^{*}$,
there exists an $(a_l, a_r)-[[nn^{*} + a_l + a_r, nn^{*}-2k_1 k_3]]$ QSC that corrects up to at least $\lfloor \frac{d-1}{2} \rfloor$
phase errors and up to at least $\lfloor \frac{d_2 d_4 -1}{2} \rfloor $ bit errors, where $d$ is the minimum distance
of the code ${(C_1 \otimes C_3)}^{\perp}$, which satisfies $d \geq d_2 d_4$.
\end{thm}

\pf Since $\gcd( n, n^{*})=1$, it follows that the product code $C_2\otimes C_4$ (consequently, $C_1 \otimes C_3$ and ${(C_1 \otimes
C_3)}^{\perp}$) is also cyclic \cite[Theorem 1, Page 570]{Macwilliams:1977}.
The elements of the code $C_1 \otimes C_3$ are linear combinations of vectors $v_{i}^{(1)} \otimes w_{j}^{(3)}$,
where $v_{i}^{(1)} \in C_1$ and $w_{j}^{(3)} \in C_3$, i.e. every $c \in C_1 \otimes C_3$ can be written as
$c=\displaystyle\sum_{i} v_{i}^{(1)} \otimes w_{i}^{(3)}$.
An (Euclidean) inner product on $C_1 \otimes C_3$ is defined as
\begin{eqnarray}\label{equation1}
\langle v_{i}^{(1)} \otimes w_{i}^{(3)} | v_{j}^{(1)} \otimes
w_{j}^{(3)}\rangle = \langle v_{i}^{(1)} | v_{j}^{(1)} \rangle
\langle w_{i}^{(3)} | w_{j}^{(3)} \rangle,
\end{eqnarray}
and it is extended by linearity for all elements of $C_1 \otimes C_3$.
Note that $\langle c_{i}^{(1)} | c_{j}^{(1)} \rangle$ and
$\langle c_{i}^{(3)} | c_{j}^{(3)} \rangle$ are the Euclidean inner
products on $C_1$ and $C_3$, respectively.
From (\ref{equation1}), since $C_1$ is self-orthogonal, $C_1 \otimes
C_3$ is also self-orthogonal, so ${(C_1 \otimes C_3)}^{\perp}$ is dual-containing.
The parameters of the codes ${(C_1 \otimes C_3)}^{\perp}$ and $C_2 \otimes C_4$
are $[nn^{*}, nn^{*} - k_1 k_3 , d]$ and $[nn^{*}, k_2 k_4 , d_2 d_4 ]$, respectively.
Since ${(C_1 \otimes C_3 )}^{\perp} \subsetneq C_2 \otimes C_4$, it follows that $d \geq d_2 d_4$.
Applying Theorem~\ref{thm:fuji} to the cyclic codes ${(C_1 \otimes
C_3)}^{\perp}$ and $C_2 \otimes C_4$, the result follows.
\qed

\begin{center}
\textbf{Acknowledgements}
\end{center}
The authors would like to thank H. Zitouni for constructing the
codes in Table~\ref{table1}. G.G. La Guardia has been partially
supported by the Brazilian Agencies CAPES and CNPq.


\begin{thebibliography}{99}

\bibitem{aly}
S.A. Aly and A. Klappenecker.
\newblock On quantum and classical BCH codes.
\newblock {\em IEEE Trans. Inform. Theory}, 53(3):1183--1188, 2007.

%\bibitem{BGG}
%A. Batoul, K. Guenda and T.A. Gulliver.
%\newblock Repeated-root isodual cyclic codes over finite fields.
%\newblock In {\em Proc. Int. Conf. on Codes Cryptology and Inform. Security},
%S. El Hajji, A. Nitaj, C. Carlet and E. Souidi (Eds.),
%LNCS 9084, Rabat, 2015.

\bibitem{repeat}
G. Castagnoli, J.L Massey, P. Schoeller and N. Von Seemann.
\newblock On repeated-root cyclic codes.
\newblock {\em IEEE Trans. Inform. Theory}, 37(2):337--342, 1991.

\bibitem{BCH-I:1960}
R.C. Bose and D.K. Ray-Chaudhuri.
\newblock Further results on error correcting binary group codes.
\newblock {\em Information and Control}, 3:279--290, 1960.

\bibitem{BCH-II:1960}
R.C. Bose and D.K. Ray-Chaudhuri.
\newblock On a class of error correcting binary group codes.
\newblock {\em Inform. Control}, 3:68--79, 1960.

\bibitem{charpin}
P. Charpin, A. Tiet\"av\"ainen and V. Zinoviev.
\newblock On binary cyclic codes with minimum distance $d = 3$.
\newblock {\em Prob. Inform. Transmission}, 33(4):287--296, 1997.

\bibitem{ezerman1}
M.F. Ezerman, S. Ling and P. Sol\'e.
\newblock Additive asymmetric quantum codes.
\newblock {\em IEEE Trans. Inform. Theory}, 57(8):5536--5550, 2011.

\bibitem{fujiwara1}
Y. Fujiwara.
\newblock Block synchronization for quantum information.
\newblock {\em Phys. Rev. A}, 87(02):23--44, 2013.
\bibitem{fujiwara2}

Y. Fujiwara, V.D. Tonchev and T.W.H. Wong.
\newblock Algebraic techniques in designing quantum synchronizable codes.
\newblock {\em Phys. Rev. A}, 88(1):012318, 2013.

\bibitem{fujiwara3}
Y. Fujiwara and P. Vandendriessche.
\newblock Quantum synchronizable codes from finite geometries.
\newblock {\em IEEE Tran. Inform. Theory}, 60(11):7345--7354, 2014.

\bibitem{G}
K. Guenda.
\newblock Quantum duadic and affine-invariant codes.
\newblock {\em Int. J. Quantum Inform.}, 7(1):373--384, 2009.

\bibitem{GGISIT}
K. Guenda and T.A. Gulliver.
\newblock Self-dual repeated root cyclic and negacyclic codes over finite fields.
\newblock In {\em Proc. IEEE Int. Symp. Inform. Theory},
2914--2918, 2012.

\bibitem{GG}
K. Guenda and T.A. Gulliver.
\newblock Symmetric and asymmetric quantum codes.
\newblock {\em Int. J. Quantum Inform.},
11(5):1350047, 2013.

\bibitem{huffman03}
W.C. Huffman and V. Pless.
\newblock {\em Fundamentals of Error-Correcting Codes}.
\newblock Cambridge Univ. Press, New York, 2003.

\bibitem{ioffe}
L. Ioffe and M. M\'ezard.
\newblock Asymmetric quantum error-correcting codes.
\newblock {\em Phys. Rev. A}, 75(3):032345, 2007.
%\bibitem{laguardia:2009}
%G. G. La Guardia.
%\newblock Constructions of new families of nonbinary quantum codes.
%\newblock {\em Phys. Rev. A}, 80(4):042331(1--11), October 2009.

\bibitem{laguardia:2012}
G.G. La Guardia.
\newblock Asymmetric quantum Reed-Solomon and generalized Reed-Solomon codes.
\newblock {\em Quantum Inform. Process.}, 11:591--604, 2012.

\bibitem{laguardia:2013}
G.G. La Guardia.
\newblock Asymmetric quantum codes: New codes from old.
\newblock {\em Quantum Inform. Process.}, 12:2771--2790, 2013.

\bibitem{lidl}
R. Lidl and H. Niederreiter.
\newblock {\em Introduction to Finite Fields and their Applications}.
\newblock Cambridge Univ. Press, Cambridge, UK, 1975.

\bibitem{Lidl:1997}
R. Lidl and H. Niederreiter.
\newblock {\em Finite Fields}.
\newblock Cambridge Univ. Press, Cambridge, UK, 1997.

\bibitem{Macwilliams:1977}
F.J. MacWilliams and N.J.A. Sloane.
\newblock {\em The Theory of Error-Correcting Codes}.
\newblock North-Holland, 1977.

\bibitem{probst}
M. Probst and L. Trieloff.
\newblock Bit and Frame Synchronization Techniques, 2003.

\bibitem{sarvepali}
P.K. Sarvepalli, A. Klappenecker and M. Rotteler.
\newblock Asymmetric quantum codes: Constructions, bounds and performance.
\newblock {\em Proc. R. Soc. A}, 465(2105):1645--1672, 2009.

\bibitem{smid}
M.H.M. Smid.
\newblock Duadic codes.
\newblock {\em IEEE. Trans. Inform. Theory},
33(3):432--433, May 1987.

\bibitem{fujiwara4}
X. Yixuan, J. Yuan and Y. Fujiwara.
\newblock Quantum synchronizable codes from quadratic residue codes and their supercodes.
\newblock In {\em Proc. IEEE Inform. Theory Workshop},
Hobart, TAS, 172--176, Nov. 2014.

\bibitem{Ge} T. Zhang and G. Ge.
\newblock Quantum block and synchronizable codes derived from certain classes of polynomials.
\newblock Available online: http://arxiv.org/abs/1508.00974.
\end{thebibliography}
\end{document}